\documentclass[conference]{IEEEtran}
\usepackage{balance}
\IEEEoverridecommandlockouts
\usepackage{adjustbox}
\usepackage{booktabs}
\usepackage{subfigure} 

\usepackage{epstopdf}
\usepackage{xcolor}
\usepackage{xspace}
\usepackage{textcomp}
\usepackage{pifont}
\usepackage{array}
\usepackage{float}
\usepackage{graphicx}
\usepackage{tabularx}
\usepackage{multirow}
\usepackage{cite}

\newif\ifdraft
\draftfalse
\drafttrue
\ifdraft
  \newcommand{\ian}[1]{{\textcolor{red}{ Ian: #1 }}}
  \newcommand{\ryan}[1]{{\textcolor{magenta}{ Ryan: #1 }}}
  \newcommand{\kyle}[1]{{\textcolor{teal}{ Kyle: #1 }}}
  \newcommand{\franck}[1]{{\textcolor{brown}{ Franck: #1 }}}
\else
  \newcommand{\ian}[1]{}
  \newcommand{\ryan}[1]{}
  \newcommand{\kyle}[1]{}
  \newcommand{\franck}[1]{}  
\fi

\usepackage{tikz}
\newcommand*\circled[1]{\tikz[baseline=(char.base)]{
            \node[shape=circle,fill,inner sep=0.5pt] (char) {\textcolor{white}{#1}};}}
            
\usepackage{amsmath,amssymb,amsfonts}
\def\BibTeX{{\rm B\kern-.05em{\sc i\kern-.025em b}\kern-.08em
    T\kern-.1667em\lower.7ex\hbox{E}\kern-.125emX}}

\begin{document}

\renewcommand\IEEEkeywordsname{Keywords}

\title{Optimizing Scientific Data Transfer on Globus with Error-bounded Lossy Compression}

\author{\IEEEauthorblockN{Yuanjian Liu\IEEEauthorrefmark{1}, Sheng Di\IEEEauthorrefmark{2}, Kyle Chard\IEEEauthorrefmark{1}\IEEEauthorrefmark{2}, Ian Foster\IEEEauthorrefmark{1}\IEEEauthorrefmark{2}, Franck Cappello\IEEEauthorrefmark{2}}
\IEEEauthorblockA{\IEEEauthorrefmark{1}
University of Chicago, Chicago, IL, USA}
\IEEEauthorblockA{\IEEEauthorrefmark{2}
Argonne National Laboratory, Lemont, IL, USA
}
yuanjian@uchicago.edu, sdi1@anl.gov, chard@uchicago.edu, foster@anl.gov,
cappello@mcs.anl.gov
\thanks{Corresponding author: Sheng Di, Mathematics and Computer Science Division, Argonne National Laboratory, 9700 Cass Avenue, Lemont, IL 60439, USA}
}

\maketitle

\thispagestyle{plain}
\pagestyle{plain}

\begin{abstract}

 The increasing volume and velocity of science data 
 necessitate the frequent movement of enormous data volumes
 as part of routine research activities. As a result, 
 limited wide-area bandwidth often leads to bottlenecks
 in research progress. However, in many cases, consuming applications (e.g., for analysis, visualization, and machine learning) can achieve acceptable performance on reduced-precision data, and thus researchers may wish to compromise on data precision to reduce transfer and storage costs.
 Error-bounded lossy compression presents a promising approach as it can 
 significantly reduce data volumes while preserving data integrity based on user-specified error bounds. In this paper, we propose a novel data transfer framework called \textit{Ocelot} that integrates error-bounded lossy compression into the Globus data transfer infrastructure. 
 We note four key contributions: (1) Ocelot is the first integration of lossy compression in Globus to significantly improve scientific data transfer performance over wide area network (WAN).  
 (2) We propose an effective machine-learning based lossy compression quality estimation model that can predict the quality of error-bounded lossy compressors, which is fundamental to ensure that transferred data are acceptable to users.
 (3) We develop optimized strategies to reduce the compression time overhead, counter the compute-node waiting time, and improve transfer speed for compressed files.
 (4) We perform evaluations using many real-world scientific applications across different domains and distributed Globus endpoints. Our experiments show that Ocelot can improve dataset transfer performance substantially, and the quality of lossy compression (time, ratio and data distortion) can be predicted accurately for the purpose of quality assurance.

\end{abstract}


\pagenumbering{arabic}
\section{Introduction}


Large amounts of data are being produced by high performance computing (HPC) simulations and advanced instruments such as the Advanced Photon Source (APS)~\cite{APSU} and LCLS-II~\cite{lcls-ii}. 
These data typically need to be shared for analysis, storage, publication, and archival, and often across multiple research institutions. 
However, transferring data over a wide area network (WAN) can be time-consuming, significantly delaying research progress.
Tools like Globus~\cite{Globus,data-transfer-between-sites} have been widely adopted to improve data transfer performance; however, while transfer performance can be increased by deploying more data transfer nodes or creating parallel data streams, limited network bandwidth constrains transfer speeds.

Many scientific data are collections of floating point numbers, and often
scientific applications do not require the level of precision encoded in those data. Thus, it is possible to reduce the data size by compromising the precision to a certain level. Error-bounded lossy compression exploits this fact and offers the potential to significantly reduce data sizes.
However, optimal tuning of compression process (i.e., for performance and quality) remains an open problem and thus such methods are rarely used
in data transfer solutions.
Although error-bounded lossy compression can substantially reduce the volume of data with user-tolerable data distortion, existing studies focus on conventional use cases such as reducing storage space \cite{mdz}, lowering I/O cost \cite{data-dumping-sz-cluster19}, or reducing memory capacity requirements \cite{quantum-compute-compression,DeepSZ}. Li et al.~\cite{resilient-data-compression} studied how to make the error-bounded lossy compressor SZ resilient to soft errors during data transfers and evaluated their approach by using a numerical analysis/simulator, but they did not systematically model and optimize data transfer performance with respect to lossy compression techniques. 

Modeling and optimizing the error-bounded lossy compression based data transfer over WAN is challenging in practice. On one hand, adding compression/decompression into the transfer services introduces new complexities (compute nodes will be involved, compressors need to be configured, overall transfer performance will be influenced by the compression speed, etc.).
On the other hand, it is critical for users to understand the quality of compressed data, so that they can precisely control the data distortion and/or meet expected compression ratios for their use cases. However, scientific applications are distinct from each other and lossy compressors exhibit different characteristics/performance because of their distinct designs. It is non-trivial to predict the compression ratios and quality accurately.

In this paper, we propose an optimized data transfer model, namely \textit{Ocelot}, by leveraging error-bounded lossy compression techniques in data transfers. 
Our contributions are:

\begin{itemize}
    \item We develop an efficient lossy compression quality prediction model, which is fundamental 
    to accurately predict the data distortion of lossy reconstructed data and compression ratio/speed. 
    \item We propose a novel approach for efficient wide-area data transfer  by combining the error-bounded lossy compression techniques, Globus~\cite{Globus:2014001,Globus:2015001,Globus}, and FuncX, a federated-Function-As-a-Service (FaaS) platform~\cite{funcX}. We also optimize the performance by developing a series of strategies to address I/O contention, compute-node waiting, and transfer slow-down for many small files. 
    \item We evaluate Ocelot using several Globus endpoints and real-world scientific applications across different domains. Experiments show that applying parallel compression can significantly improve data transfer performance over WAN (reaching 11.2$\times$ speed-up with negligible data distortion for users). 
\end{itemize}


The rest of the paper is organized as follows. In Section \ref{sec:related}, we discuss related work. In Section \ref{sec:background}, we present the research background. In Section \ref{sec:formulation}, we propose the online data transfer framework Ocelot, which integrates error-bounded lossy compression technology with Globus. 
In Section \ref{sec:usecases}, we describe three capabilities of Ocelot. In Sections \ref{sec:predict_model} and \ref{sec:performance-optimization}, we describe 
how we conduct lossy compression quality prediction and optimize  data transfer performance with lossy compression techniques, respectively. In Section \ref{sec:evaluation}, we evaluate Ocelot on real-world scientific datasets and the state-of-the-art lossy compressor SZ with different compression pipelines.
Finally, we conclude the paper with a discussion of future work in Section \ref{sec:conclusion}.

\section{Related Work}
\label{sec:related}

In this section, we discuss the related works in two facets: the modern techniques in wide area data transfer and common use-cases of error-bounded lossy compression.

Many systems have been developed to improve the performance of large wide-area data transfers. One common method adopted by many commercial data transfer tools, such as FileCatalyst \cite{filecatalyst} and IBM Aspera \cite{ibmaspera}, is using User Datagram Protocol (UDP) or multiple Transmission Control Protocol (TCP) streams. 
BitTorrent is a popular Peer-to-Peer (P2P) data transfer software developed at the application level over TCP/IP, which can be used to transfer big data files. BitTorrent adopts a tracker/seed mechanism to allow each data downloader to be a data uploader in a community, such that the more the users participate, the higher the data transfer speed. The BitTorrent technique, however, is unsuitable for big data transfer in the science community because a stable and secure science data-sharing service is highly required. 
Globus is a research data management platform that enables high-performance, secure, and reliable third-party data transfers. 
Globus builds upon the GridFTP protocol for data movement and adopts several optimization techniques such as parallel streams \cite{end-to-end-transfer-tcp,1015527,gridftp-opt}, which can significantly improve data transfer performance. Transferring big data files with Globus, however, may still suffer from low performance in practice, as performance is related to multiple sophisticated factors such as the settings on Globus connect server (GCS) endpoints (concurrency, pipelining, striping, etc.) \cite{data-transfer-between-sites}, low quality network paths, and underprovisioned data transfer nodes (DTNs). In particular, recent studies \cite{data-transfer-between-sites} show that transferring big data files between Argonne Leadership Computing Facility (ALCF) and National Energy Research Scientific Computing Center (NERSC) could be slow (only hundreds of MB/s) at an inefficient concurrency setting.  

Error-bounded lossy compression has been effective in significantly reducing data volumes for many use cases. However, it has rarely been used in the wide area data transfer case. Common use cases for error-bounded lossy compression include reducing storage footprint \cite{n-body-compression,mdz,9086223}, reducing memory capacity requirements \cite{quantum-compute-compression,quantum-compute-compression2}, mitigating I/O costs in supercomputers \cite{data-dumping-sz-cluster19}, and avoiding recomputation of data \cite{pastri}. Zhao et al.~\cite{mdz}, for instance, developed an efficient lossy compressor for molecular dynamics (MD) simulation data based on the spatio-temporal patterns of MD datasets, which aims to reduce the storage space as much as possible. Wu et al.~\cite{quantum-compute-compression} explored the best-qualified error-bounded lossy compression method for Intel-QS \cite{smelyanskiy2016qhipster}---a full-state quantum circuit simulator developed by Intel, in order to significantly lower the required memory capacity for large-scale quantum computing simulations. Li et al.~\cite{resilient-data-compression} proposed a resilient error-bounded lossy compression method, which aims to protect the data compression against potential errors such as SDCs. However, their work does not involve data transfer performance optimization, which is instead addressed in our work. 
\section{Research Background}
\label{sec:background}

We briefly describe the critical technical components on which we build. 

\subsection{Error-bounded Lossy Compression}
\label{sec:compressors}

Error-bounded lossy compression has been broadly used to significantly reduce the volumes of scientific datasets produced by large-scale HPC applications or advanced instruments (with a compression ratio of several hundreds or thousands \cite{use-case}), while effectively controlling data distortion based on the user-specified error bound. In comparison with lossy compression, lossless compression suffers from low compression ratios ($\leq$2 in most cases \cite{lossless-scientific-data,mdz,Kai-ICDE2021}) since lossless compressors generally depend on the exactly repeated byte stream patterns while scientific datasets are often composed of floating-point data values often with diverse ending mantissa bits. 

There have been many error-bounded lossy compressors developed. In general, there are two models for error-bounded lossy compression: the transform-based model and the prediction-based model. The former performs the (near)orthogonal transform to decorrelate the raw data to another coefficient data (such as by wavelet transform) and then reduce the coefficient data by specific encoders such as embedded encoding \cite{zfp}. The typical examples are ZFP \cite{zfp} and SSEM \cite{ssem}. The latter uses a data predictor and  linear-scale quantization to decorrelate the datasets and then uses a variable-length encoding (such as Huffman encoding \cite{huffman}) and dictionary encoding (such as LZ77 \cite{lz77}) to obtain a fairly high compression ratio. Examples include SZ \cite{Kai-ICDE2021,Liu-aesz} and MGARDx \cite{liang2020mgard}. 

We adopt SZ3 \cite{sz3} in our work for two reasons: its modular structure, which allows us to construct many different compression pipelines (i.e., different compressors) for different datasets and use cases, and the high performance \cite{Kai-ICDE2021} of its default SZ-interp compression algorithm, which exhibits the highest compression ratio and quality in many cases compared with the other state-of-the-art lossy compressors including ZFP, SZ2, and MGARDx.  

\subsection{Globus Data Transfer Infrastructure}

Globus 
is a research data management platform that is used to transfer, synchronize, and share large volumes of data. 
Globus was launched in 2010, and has since managed the reliable
movement of almost two exabytes of data across 40,000 endpoints
distributed around the world. 
Globus endpoints are widely deployed at universities, research laboratories, and on cloud platforms (such as Amazon S3 and Google drive). 

Globus adopts the GridFTP protocol \cite{gridftp,gridftp-opt} to provide high-performance, secure, and reliable data transfer over WAN. There are many optimization strategies in GridFTP for improving data transfer performance, such as pipelining, parallelism, and concurrency. GridFTP pipelining avoids blocking/waiting on transfer-commands, which can transfer many small files very efficiently. Parallelism allows different portions of the same file to be sent by multiple channels in parallel. Concurrency supports transferring of different data files through multiple channels in parallel.  

\subsection{Federated Function as A Service (FuncX)}

FuncX~\cite{funcX} is a distributed and scalable function execution platform. FuncX differs from traditional cloud-hosted FaaS platforms in that it combines a centralized cloud-hosted service with a collection of user-deployed and managed endpoints. Users can deploy their own endpoints on their own resources via a small Python endpoint software. They may configure that endpoint to provision resources dynamically from various backend resource providers (e.g., batch schedulers, Kubernetes clusters, cloud instances). 
Users may register and execute Python functions in a similar way to cloud-hosted FaaS, by providing the function body and input arguments. However, unliek centralized FaaS they must also select an endpoint on which to execute that function. The FuncX service relies on an OAuth-based identity and access management platform, Globus Auth \cite{globus-auth}, to securely execute functions. FuncX leverages containers to package function codes and resolve dependencies on endpoints, and also enables multiple optimization strategies to obtain the best performance in the remote function calls, such as container warming (avoiding/reducing the container instantiation cost), executor/user batching (amortizing costs across many function requests), and prefetching (advertising the anticipated capacity to interleave network communication with computation).
\section{Ocelot: Online Data Transfer with Error-bounded Lossy Compression}
\label{sec:formulation}

\begin{figure}[ht]\centering
\includegraphics[width=0.95\columnwidth]{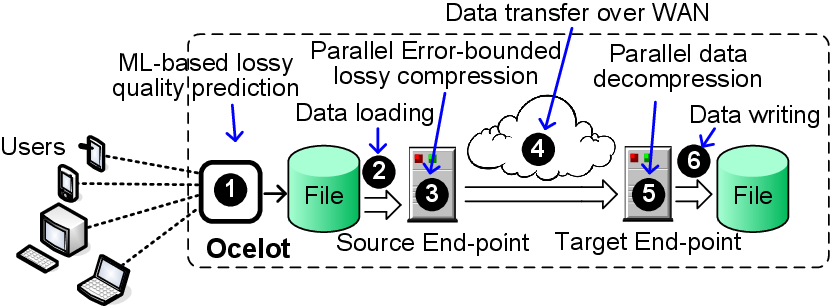}
\vspace{-3mm}
\caption{Design overview: (1) Use the quality prediction model to obtain an appropriate compressor setting; (2) Load data of various formats into the compression program; (3) One or more compute nodes on the source machine are used to compress the datasets; (4) transfer the data over WAN with Globus; (5) One or more compute nodes on the target machine are used to decompress the datasets; (6) The decompressed files are written into the disks on the target machine.}
\label{fig:design}
\end{figure}

\figurename~\ref{fig:design} presents a high-level overview of Ocelot. As shown in the figure, Ocelot provides an ML-based quality prediction model for users to predict the lossy compression quality (such as data distortion and compression ratio), thus guaranteeing the integrity/validity of the lossy reconstructed data (step \circled{1}). The data then progress through five steps (\circled{2}-\circled{6}) during the data transfer procedure from one endpoint to another over WAN. The key difference between Ocelot and the traditional data transfer method is that we integrate an error-bounded lossy compression step, which is expected to significantly reduce the data volume before transferring the data. At the target endpoint, upon receiving of compressed data, they are be decompressed and then written to the file system. The detailed compression technologies have been discussed in Section \ref{sec:compressors}. 
Ocelot can be used remotely without needing to 
manually log in to the source or destination machine to perform the compression/decompression task, because the executors have been deployed on those machines beforehand.

\begin{figure}[htb]\centering
\includegraphics[width=0.8\linewidth]{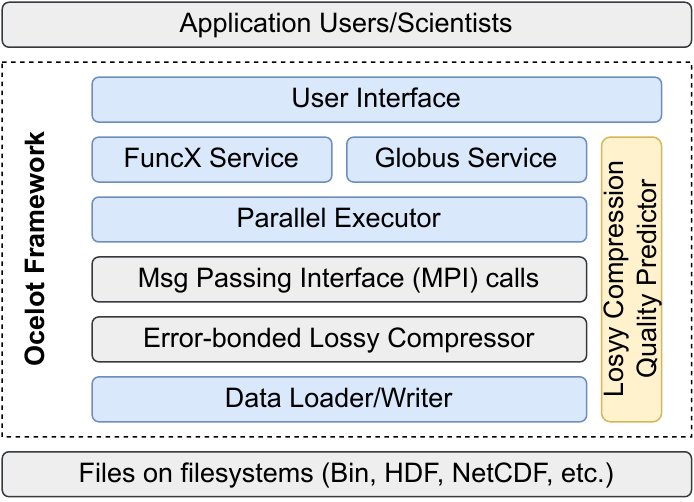}
\vspace{-3mm}
\caption{System architecture: Six new modules (colored) are added to form the Ocelot framework: (1) Data loader can load data of multiple formats including NetCDF, HDF5\cite{hdf5}, binary; (2) The lossy compression quality predictor is used to find a suitable error bound and compressor to conduct the compression; (3) Parallel executor handles the compression/decompression tasks; (4) FuncX service deals with remote orchestration; (5) Globus manages the data transfer; (6) User interface offers a graphical interface that helps users submit the tasks easily.}
\label{fig:architecture}
\end{figure}

We present our architecture in \figurename~\ref{fig:architecture} (the colored boxes indicate the new modules we developed for Ocelot). In our design, the user connects the Ocelot Framework through a user interface (e.g., a command line or GUI). Upon receiving user's data transfer task, Ocelot starts the quality predictor via funcX to
obtain a suitable compressor configuration by testing a few settings very quickly with subsampling methods. funcX allows these tasks to be executed on the remote resource on which the data reside. Ocelot then uses funcX to  initiate a compression task on the remote endpoint. 
The compression is conducted by an MPI program that loads different files from the file systems and compresses them in parallel. Ocelot then
starts the transfer via Globus. The transfer will move the compressed files to the target machine once the files are ready. There is some optimization here because sometimes the compression tasks cannot be scheduled immediately. We will leave the detailed discussion to Section \ref{sec:performance-optimization}.
We design Ocelot to be flexible, enabling users to bypass the quality predictor by manually providing a compressor configuration for certain cases when they know what error bound and compressor to use. The quality predictor module is driven by our designed machine-learning model, which will be detailed in Section \ref{sec:predict_model}.

 
\section{Critical capabilities offered by Ocelot}
\label{sec:usecases}
Before diving into the technical details, we introduce three key capabilities of our framework from a user's perspective.
\begin{enumerate}
    \item Selecting best-qualified lossy compression configuration based on our proposed quality predictor.
    
    Ocelot is able to select the most suitable lossy compression configuration in terms of users' requirements. Based on the estimated results generated by our quality predictor, the user can select the ``best'' compression solution for their data. Specifically, users can view the data distortion, compression ratio, and compression time for different lossy compression pipelines or configurations, thus guiding them in selecting/optimizing the best-qualified setting.
    
    \item Reducing transfer time with parallel (de)compression.
    
    After applying the prediction model to configure compression automatically, users can utilize  Ocelot to reduce the file transfer time. 
    Users need only to specify data paths and start the transfer. 
    The compression/decompression will be performed automatically.
    \item Remote orchestrating (de)compression and transfer.
    
    We incorporate FuncX and Globus Transfer API into our framework, allowing users to control the compression and transfer between endpoints on any authorized machines. Users do not have to explicitly connect to remote resources (e.g., via ssh) to submit batch jobs to do compression. Instead, they just need to run our Ocelot software on their personal computer and control the compression, transfer, and decompression remotely. 
    Moreover, Ocelot allows users to collect information about compression and transfer. The analytical data is stored on the user's personal computer, and can be used to further analyze the performance with graphical tools.
\end{enumerate}







\section{Compression Quality Prediction}
\label{sec:predict_model}

In this section, we propose a prediction model to estimate the lossy compression ratio, compression speed, and peak-to-noise ratio (PSNR) \cite{z-checker}. 
In general, 
it is impossible for users to predict compression quality (such as compression ratio and data distortion level) for a particular error-bounded lossy compressor without performing the compression on the given dataset. This is because the effect of data prediction/transform and coding in the compressor varies with diverse data features.  
With our prediction model, users can quickly test multiple compression settings and choose the one that best matches their use case.
\begin{figure}[htb]
    \centering
    \includegraphics[width=1\linewidth]{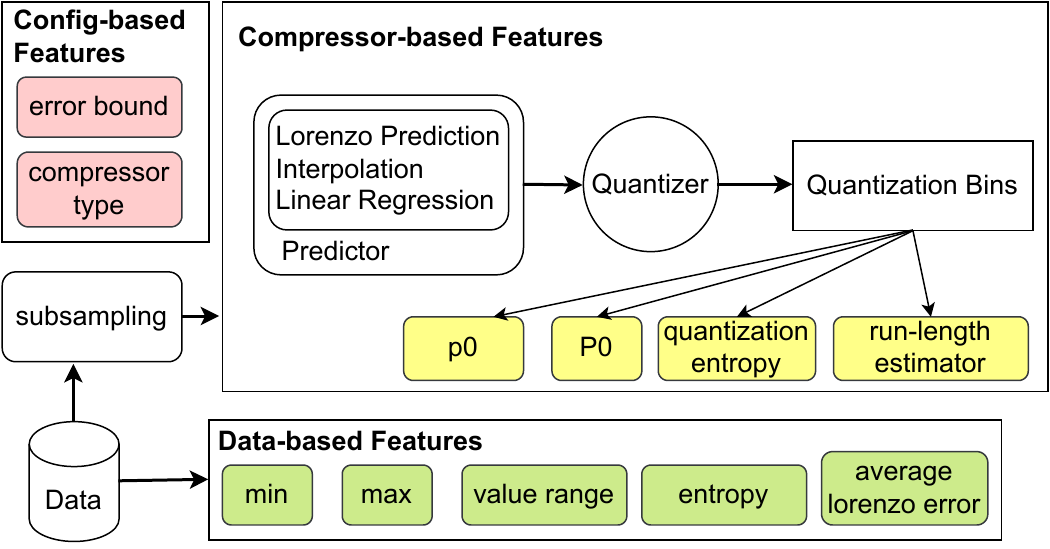}
    \vspace{-8mm}
    \caption{The features used to predict compression quality are categorized into three types: config-based, compressor-based, and data-based features, which are shown as colored boxes.}
    \label{fig:prediction-model}
\end{figure}

We train a machine learning (ML) model on masses of sample datasets, with the aim to  build a relationship between the compression-related features and the compression quality. 
The model can then be used to estimate  compression quality accurately based on the features extracted from the given datasets at runtime.  

We derive many features as input to our model,
as illustrated in  \figurename~\ref{fig:prediction-model}. Identifying a set of useful features is challenging, because (1) the extraction of each feature should have low computation cost, and (2) the features should form an accurate indicator of the compression quality. We consider features in one of three categories: (1) config-level features, (2) data-based features, and (3) compressor-level features. 

\textbf{Config-based features} are configuration settings (including error bound values and compression pipeline) specified by users. Different error bounds can yield largely different compression quality (e.g., compression ratios and compression speed). Compression quality also depends on specific compressors each with distinct designs. The prediction-based compressors\cite{sz2,sz3}, for example, may adopt various predictors which may exhibit different performances. 
We enable our model to recognize the characteristics of compressors by treating the compressor-type feature as a discrete classification variable and feeding it with profiling data. 

\begin{figure}[ht]
    \centering
    \includegraphics[width=1\linewidth]{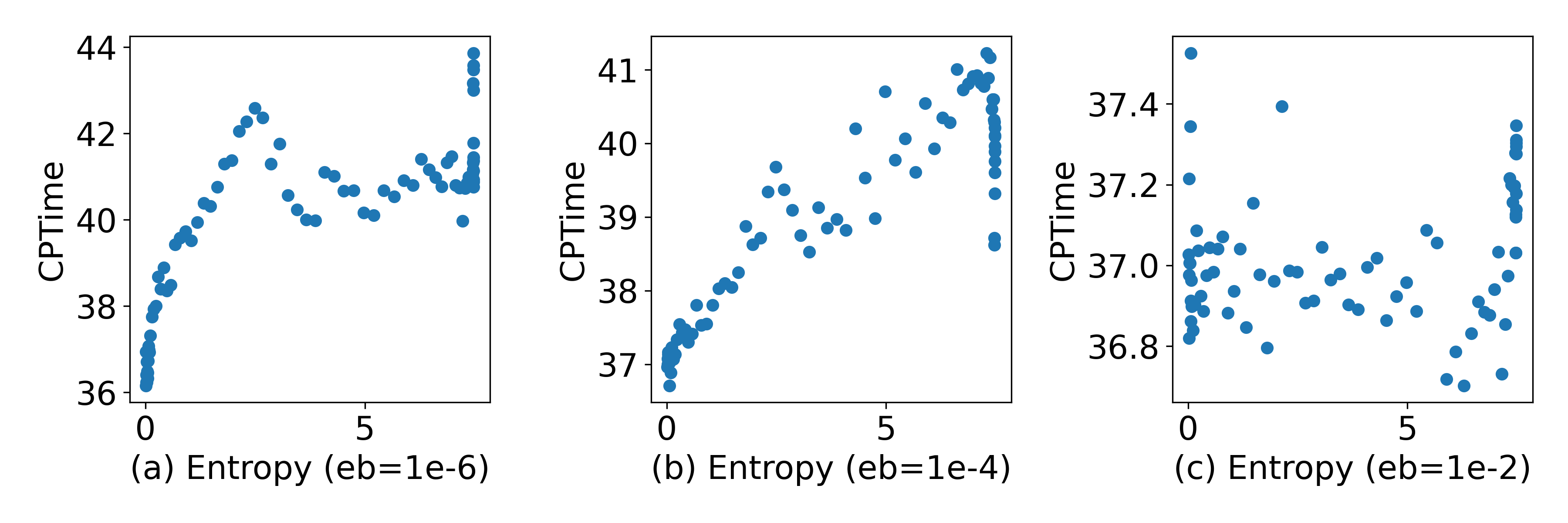}
    \vspace{-8mm}
    \caption{Data entropy vs compression time in Reverse Time Migration (RTM) \cite{rtm} application with three error bound settings}
    \label{fig:entropy-vs-cptime-rtm}
\end{figure}

\textbf{Data-based features} describe the characteristics of datasets, which is also a key factor to distinguish the compressibility. As shown in \tablename~\ref{tab:data-level-features}, even for the same application, different datasets can have very different properties such as min, max, and value range. In addition, we also use byte-level information entropy as one feature, because it reflects the ``chaos-level'' of a dataset. The entropy is defined as
$$H(X) = - \sum\nolimits_{x \in S}{p(x)\log p(x)} = E[-\log p(X)]$$
where $S$ is the set of byte values (0-255) and $p$ denotes the probability/frequency of an element in $S$.  
In general, the higher entropy a dataset exhibits, the more difficult it is to compress that dataset. As verified in \figurename~\ref{fig:entropy-vs-cptime-rtm} (a) and (b), the entropy value projects a positive correlation against the compression time, especially when the error bound is relatively low. It is worth noting that when the error bound is relatively high, the entropy would lose its effect (as shown in \figurename~\ref{fig:entropy-vs-cptime-rtm} (c)), because the large error bound would diminish the data variation. Moreover, we use the average Lorenzo error (i.e., the difference between the true data value and Lorenzo-predicted value\cite{sz2}) as a feature to shape the ``easiness of prediction'' for a dataset. If the average Lorenzo error is high, the prediction stage tend to be imprecise, leading to low compression ratio.

\begin{table}[htb]
\centering
\caption{Examples of the basic data-based features in different datasets: CLDHGH, FLDSC, PCONVT are three fields in the CESM\cite{cesm} dataset. HACC-VX and HACC-VY are two fields in the HACC\cite{hacc} dataset.}
\resizebox{\linewidth}{!}{
\begin{tabular}{|l|l|l|l|l|l|}
\hline
\textbf{Dataset} & \textbf{CLDHGH} & \textbf{FLDSC} & \textbf{PCONVT} & \textbf{HACC-VX} & \textbf{HACC-XX} \\ \hline
min              & 0.00            & 92.84          & 39025.27        & -3846.21         & 0.00             \\ \hline
max              & 0.92            & 418.24         & 103207.45       & 4031.25          & 256.00           \\ \hline
value range      & 0.92            & 325.40         & 64182.18        & 7877.46          & 256.00           \\ \hline
\end{tabular}
\label{tab:data-level-features}}
\end{table}

\textbf{Compressor-based features} are the properties of the intermediate data used in the course of lossy compression, which generally have the highest prediction ability for compression quality. Specifically, we focus on the quantization bins, as shown in \figurename~\ref{fig:prediction-model}. Since the quantization bins are encoded by the subsequent lossless encoders, its characteristic closely correlates to the final compression quality. In order to control the execution overhead, the quantization bins are computed based on the sampled data points. As demonstrated in  \figurename~\ref{fig:prediction-model}, we develop four compressor-based features, including $p_0$, $P_0$, quantization entropy, and run-length estimator. (1) $p_0$ denotes the percentage of the 0-value bins over all quantization bins. 
In general, large $p_0$ tends to yield a high compression ratio and compression speed, because a large majority of predictions should be accurate in this situation. (2) $P_0$ denotes the fraction of `0'(encoded) taken in Huffman coding in the regard of the full Huffman encoded data size. (3) Quantization entropy is the entropy of quantization bins. If the prediction is accurate, quantization bin values will mostly be near 0, and the quantization entropy will be low.  (4) Run-length estimator (denoted $R_{rle}$) is derived from $P_0$ and $p_0$ by the following equation:
$R_{rle} = 1/((1-p_0)P_0 + (1-P_0))$. 



\begin{figure}[htb]
    \centering
    \includegraphics[width=1\linewidth]{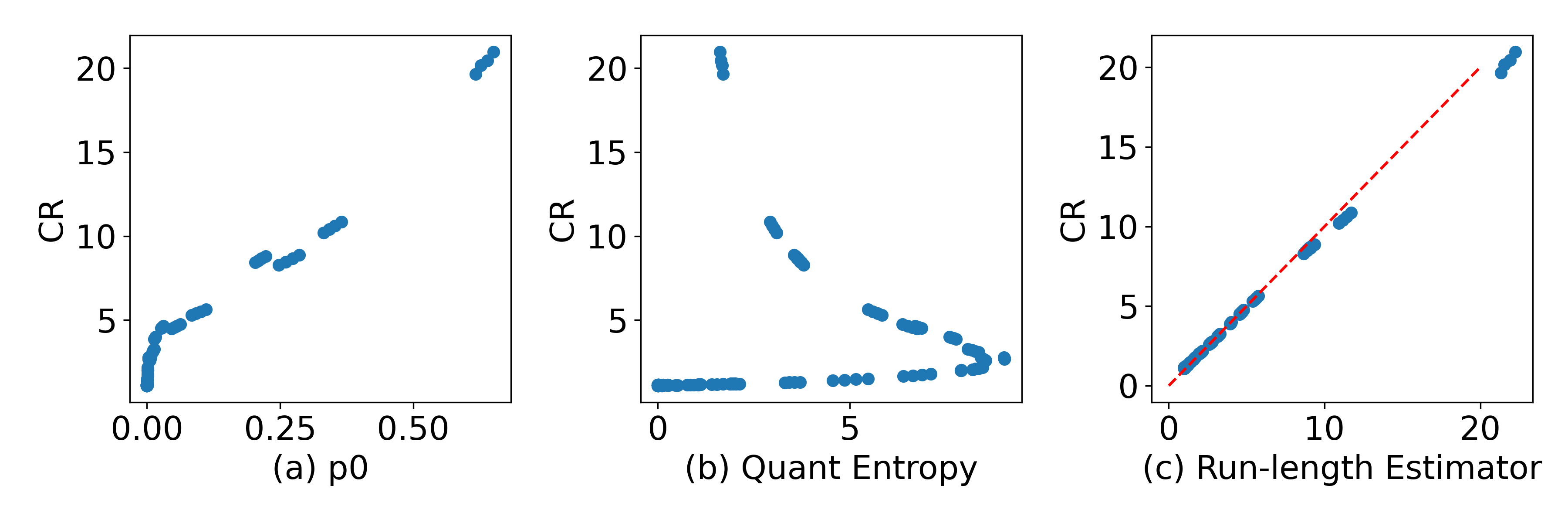}
    \vspace{-8mm}
    \caption{The relationship between $p_0$, quantization entropy, run-length estimator and compression ratio for Nyx application.}
    \label{fig:NYX-CR}
\end{figure}
\begin{figure}[htb]
    \centering
    \includegraphics[width=1\linewidth]{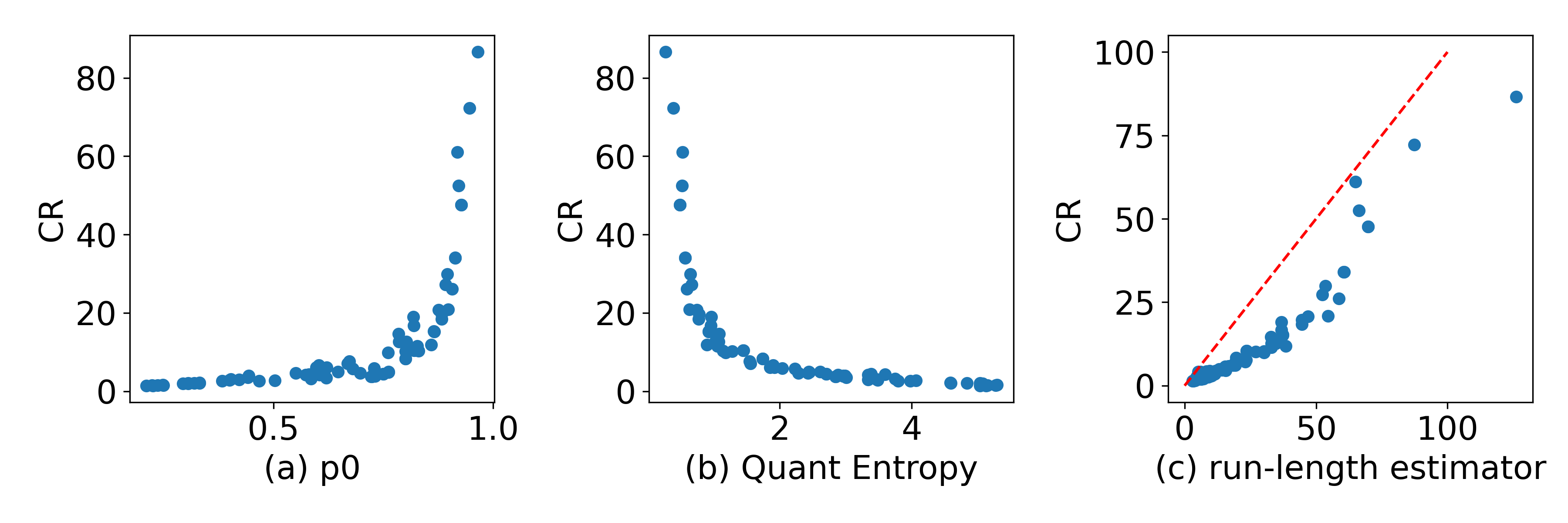}
    \vspace{-8mm}
    \caption{Run-length estimator alone fails to predict the compression ratio for Miranda application while the three features together form a correlation to the compression ratio which can be learned by a machine-learning model.}
    \label{fig:miranda-cr}
\end{figure}

Although the $p_0$ and $P_0$ are also used in related work \cite{sian-ratio}, our solution is much more accurate in compression quality estimation in general cases. 
The estimation of compression ratio in \cite{sian-ratio} depends on the following formula: $\hat{CR}=1/(C_1(1-p_0)P_0 + (1-P_0))$, where $C_1$ is an ad-hoc tuning parameter which varies with different applications. As shown in \figurename~\ref{fig:NYX-CR} (c), almost all data points are located on the line $y=x$ (red line in the figure), which means the estimated compression ratio $\hat{CR}$ under that formula could be very accurate in this case. This is due to the fact that this formula happens to form a linear function with compression ratio for the Nyx\cite{nyx} application.
However, that formula is sensitive to the tuning of the $C_1$ parameter, which may cause unexpected large compression quality estimation errors in other applications. For instance, the estimator's value does not form a linear relationship with the compression ratio for the Miranda\cite{miranda} application (as shown in \figurename~\ref{fig:miranda-cr} (a) and (b)), which leads to bad compression quality estimation in turn (see \figurename~\ref{fig:miranda-cr} (c)). 
In comparison, our $R_{rle}$ formula does not depend on the $C_1$. In fact, $R_{rle}$ serves as a feature and we feed it into the ML model along with other features (including $p_0$ and $P_0$), and thus the model can automatically fine-tune the coefficients applied on those features, thus being able to keep an accurate estimation in most of cases (to be shown later).

\begin{figure}[htb]
    \centering
    \includegraphics[width=1\linewidth]{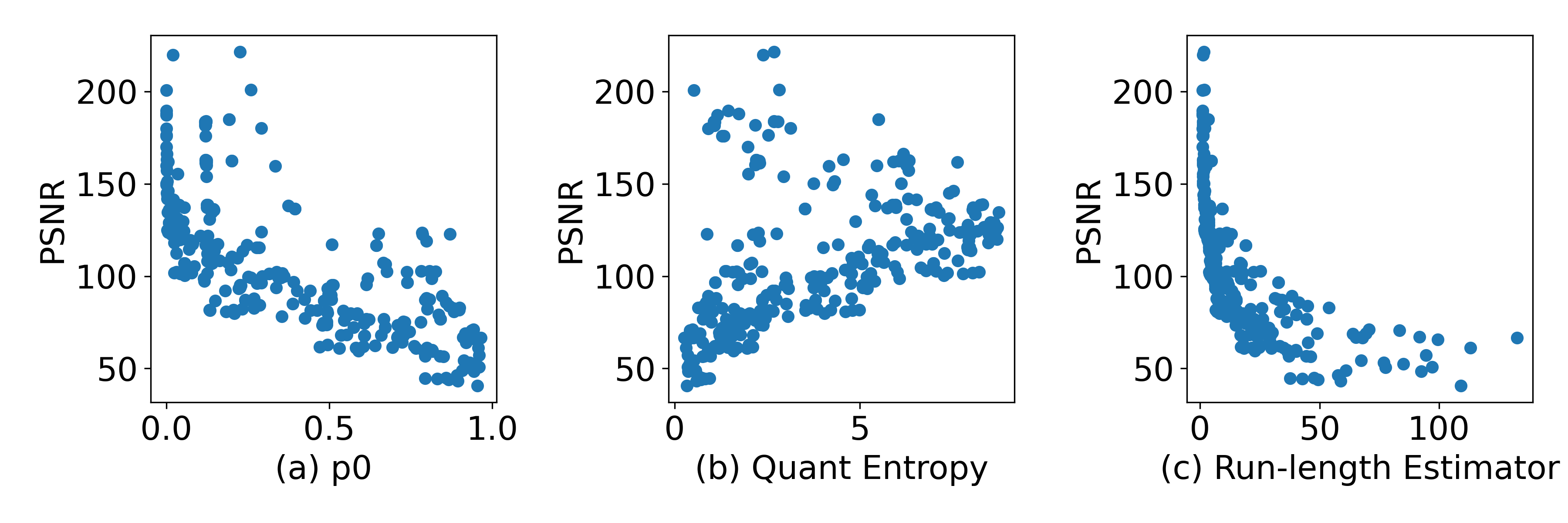}
    \vspace{-8mm}
    \caption{CESM dataset --- PSNR versus compressor-level features}
    \label{fig:CESM-PSNR}
\end{figure}

\begin{figure}[htb]
    \centering
    \includegraphics[width=1\linewidth]{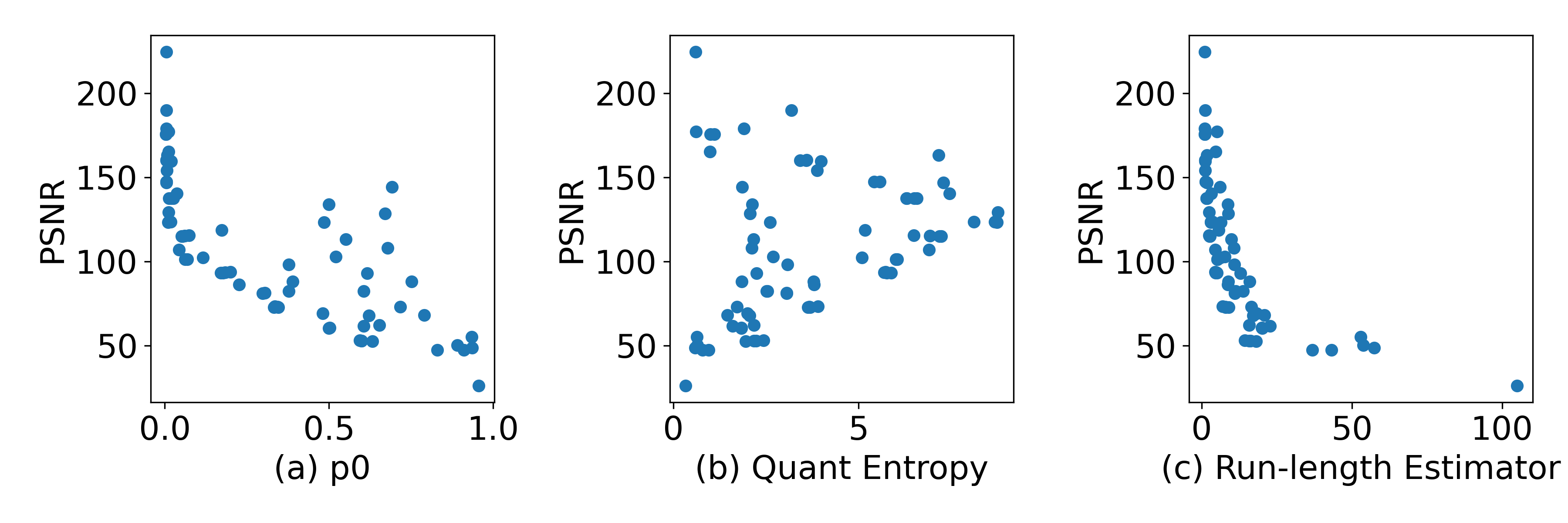}
    \vspace{-8mm}
    \caption{ISABEL dataset --- PSNR versus compressor-level features}
    \label{fig:ISABEL-PSNR}
\end{figure}

Our compressor-based features can also be used to predict the reconstructed data distortion. This is because these features are also closely correlated to the data distortion metrics such as PSNR, as verified in \figurename~\ref{fig:CESM-PSNR} and \figurename~\ref{fig:ISABEL-PSNR}. 

Based on the observations above, we use a decision tree model to perform the compression quality estimation. The evaluation result will be demonstrated in Section \ref{sec:evaluation}.
\section{Optimization of Data Transfer with Error-bounded Lossy Compression}
\label{sec:performance-optimization}
The compression performance prediction model described above provides a fast and automatic way to determine appropriate compressor settings. However, compression remains a computational expensive process, especially with large data. In Ocelot we 
utilize multiple cores/nodes to compress files in parallel.

Nonetheless, it is worth noting that there are two issues that may impede the ``compress and transfer'' performance. 
First, for large datasets the compression task may exceed the capacity available on DTNs or login nodes, and thus require provisioning of compute nodes via batch scheduler. Such requests may not be scheduled immediately. 
Second, the number and size of files significantly influence the transfer speed because (1) each file transfer has an inevitable data handling cost in addition to data transfer time, and many small files may significantly lower the overall transfer throughput; (2) transfers with too few files cannot utilize the available concurrent transfer threads.

We describe our transfer performance optimization strategies in this section. To address the first issue, we need a strategy to transfer files when compute nodes are not immediately available. For the second issue, we need an efficient file grouping method to counter issues with many compressed small files. 

\subsection{Parallel Compression/Decompression}

\begin{figure}[htb]
    \centering
    \includegraphics[width=1\linewidth]{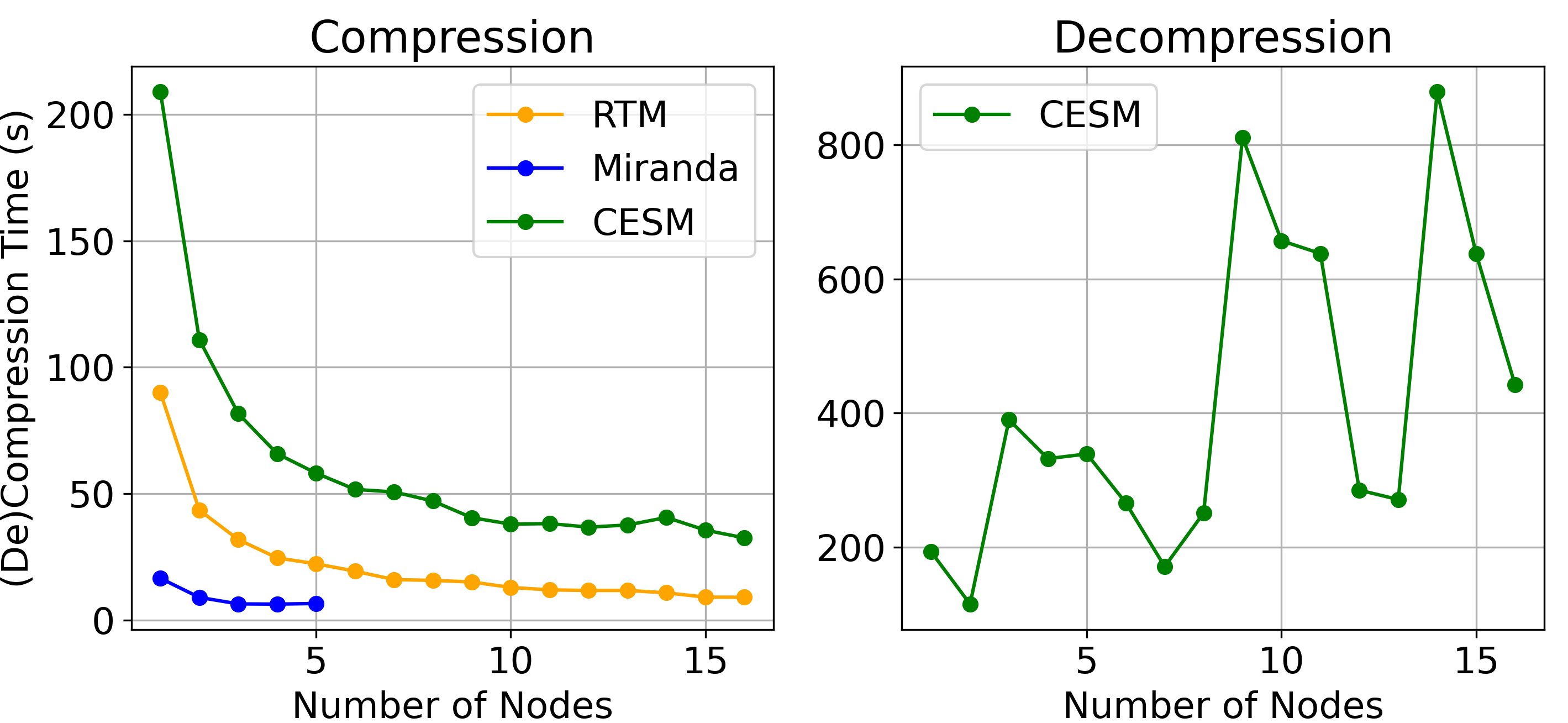}
    
    \vspace{-2mm}
    
    \caption{Parallel compression and decompression times vs.\ number of nodes, as measured on Purdue Anvil. Each node has 128 CPU cores.}
    \label{fig:parallel-compression-anvil}
\end{figure}

Our fundamental approach to reduce the transfer time is using compression methods to reduce the file sizes. However, each compression suffers 
a certain time cost, thus if we compress thousands of files sequentially, the total compression time may surpass the transfer time. 
We utilize parallel computing to significantly accelerate the compression process. We investigate the performance of different levels of parallelization: as shown in \figurename~\ref{fig:parallel-compression-anvil} (left), the increase in the number of CPU cores significantly reduces the time needed to compress these datasets because they consist of many independent files. To address this issue, we let each core handle the compression of a set of files in parallel. The compression time cannot be further reduced when the number of cores reaches the number of files to be compressed because of the saturation of the parallelism.

Our experiments show that decompression performance does not increase monotonically with the number of CPU cores. For instance, decompressing the CESM~\cite{cesm} dataset on Cori takes 68.7s on four nodes but more than 5 minutes on 16 nodes.
We conduct a more thorough test for parallel decompression on the Purdue Anvil machine, and the result is shown in \figurename~\ref{fig:parallel-compression-anvil} (right). We see in this experiment that performance
degrades with more nodes. We believe this to be due to I/O contention
on a shared file system. 
We can avoid the slow-down by tuning the number of cores to the parallel file system. 

\subsection{Optimization for Node Waiting Time}

The uncertain wait time on compression tasks and transfer tasks may degrade transfer performance when involving compression. 
In most systems there are infrequently sufficient nodes available immediately to do the compression when users submit the data compression tasks. If the compression tasks are stuck too long in the scheduler queue, the overall transfer performance would be even slower than transferring without compression. 

\begin{figure}[htb]
    \centering
    \includegraphics[width=1\linewidth]{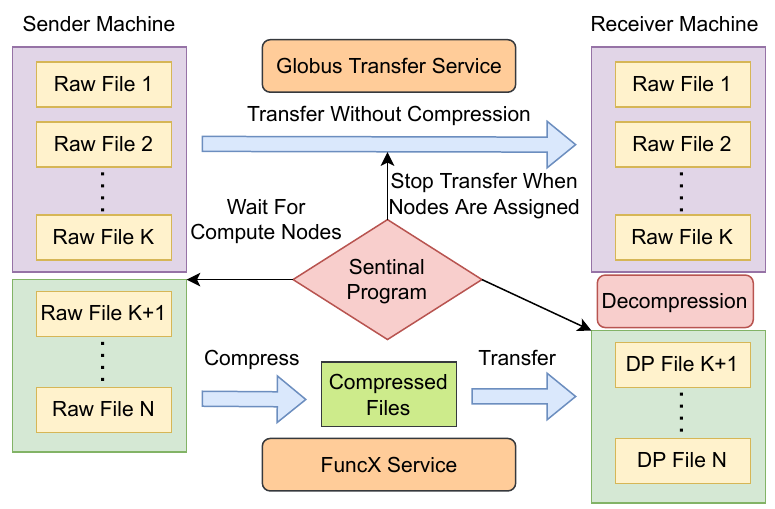}
    \vspace{-5mm}
    \caption{Transfer without compression during node waiting time: the monitoring program submit the compression task; while waiting for nodes, the transfer service is already transferring the data without compression.}
    \label{fig:node-waiting-problem}
\end{figure}

In order to counter the node waiting time, we run a sentinel program to monitor and schedule the transfer/compression task. As shown in \figurename~\ref{fig:node-waiting-problem}, when a user submits a transfer request (with lossy compression option turned on) which is not assigned compute nodes immediately, we start transferring the files in groups without compression. Once a file transfer is complete, we write their filenames in a meta file so that the compression scheduler knows which files no longer need compression. When the compute nodes are assigned, the sentinel program notifies the transfer tasks to stop and let the parallel compression scheduler take over the remaining files. In this way, the data transfer is not be suspended because of waiting for nodes, and the worst-case is that all data are transferred without compression (when the nodes are not assigned through the whole period). 
In production deployments, we anticipate that the Ocelot service could be deployed on dedicated cluster nodes (e.g., DTNs) with the approval of system administrator (similar to Globus service). In this case, wait time would be only dependent on other Ocelot transfers sharing those resources.

\subsection{File Grouping for High Data Transfer Throughput}

We propose a file grouping strategy to improve the data transfer throughput based on our observation that the number of files and file sizes may significantly affect the transfer speed (as shown in \tablename~\ref{tab:file-transfer-pattern}). Although the effective transfer speed fluctuates 
due to network and I/O contention, we generally see that the effective network speed decreases as the number of files increases, when transferring the same amount of data. 
This motivates us to optimize the file transfer speed by grouping small files together.

\begin{table}[htb]
\centering
\caption{File Transfer Patterns between two supercomputers: Nersc Cori and Argonne Bebop}
\resizebox{\linewidth}{!}{
\begin{tabular}{|l|r|r|r|r|}
\hline
\textbf{Total size} & \textbf{File size} & \textbf{\# Files} & \textbf{Speed (MB/s)} & \textbf{Duration (s)} \\ \hline
300GB               & 1M                & 300000               & 247.0               & 1235                 \\ \hline
300GB               & 10M               & 30000                & 921.1               & 325                  \\ \hline
300GB               & 100M              & 3000                 & 1120.0                 & 267                  \\ \hline
300GB               & 1000M             & 300                  & 1060.0                 & 281                  \\ \hline
\end{tabular}}
\label{tab:file-transfer-pattern}
\end{table}

\begin{figure}[htb]
    \centering
 \includegraphics[width=1\linewidth]{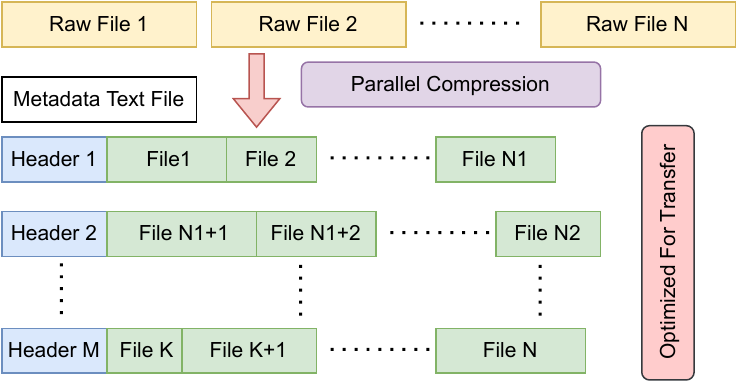}
    \caption{Parallel compression optimization by grouping small compressed files to achieve higher transfer speed.}
    \label{fig:parallel-compression-groups}
\end{figure}

Grouping small compressed files can increase a single file's size and reduce the number of files, and thus improve transfer speed. As shown in \figurename~\ref{fig:parallel-compression-groups}, we compress files in parallel and group many compressed files to achieve a better size for transfer. We use MPI to communicate the compressed sizes among CPU cores to determine the file offset for each core to write. Each grouped file has a header and a body of connected compressed data. The header contains information about the number of compressed files in this group, the starting offset, and the size of each file. The metadata text file contains human-readable information about the number of files, grouping strategy, and the original filenames that are useful for decompression. The default strategy is to group files by the ``world\_size'', i.e., the available number of cores for compression, because they run in parallel and can usually finish the compression at a similar time. According to the profiling test and information provided by the administrator, we know in advance the preferred size for each file to achieve the fastest transfer speed. Thus, the compression scheduler can also determine the number of files to put in one group based on the file sizes.

\section{Performance Evaluation}
\label{sec:evaluation}
In this section, we present our experimental testbed and performance evaluation results of our models with an in-depth analysis. We first evaluate the prediction precision on individual files with different settings and then evaluate the performance of transfer with parallel compression.

\subsection{Experimental Settings}

We collect performance data on three supercomputers: Bebop, NERSC Cori, and Purdue Anvil, with specs shown in \tablename~\ref{tab:bebop}. Each is located in different regions of the United States and has different network conditions. The evaluation of network transfer performance is based on the network connecting these supercomputers. We evaluate our prediction approaches on datasets generated by six scientific applications: QMCPACK~\cite{qmcpack}, RTM~\cite{rtm}, Miranda~\cite{miranda}, CESM~\cite{cesm}, Nyx~\cite{nyx}, and Hurricane Isabel~\cite{hurricane}, as presented in Table~\ref{tab:dataset}.

\begin{table}[htb]
\centering
\caption{Machine Specifications: bdwall and knlall are from Bebop, wholenode is from Anvil, and haswell is from NERSC cori}
\begin{adjustbox}{width=\columnwidth}
\begin{tabular}{|l|l|l|l|l|}
\hline
\textbf{Partition} & \textbf{\# Nodes} & \textbf{CPU}         & \textbf{Cores} & \textbf{Memory} \\ \hline
Bebop bdwall             & 664                & Intel Xeon  E5-2695v4 & 36             & 128GB      \\ \hline
Bebop knlall             & 348                & Intel Xeon  Phi 7230  & 64             & 96GB       \\ \hline
Anvil wholenode             & 750                & 	Two Milan CPUs @ 2.45GHz  &   128 & 256GB       \\ \hline
Cori haswell             & 2388                & 	Intel Xeon Processor E5-2698 v3  & 128 &  128GB       \\ \hline
\end{tabular}
\end{adjustbox}
\label{tab:bebop}
\end{table}

The Miranda, CESM, and RTM applications contain many files and are well-suitable for our parallel compression tasks. Specifically, we use a fixed subset of these three applications in our parallel compression evaluation. Miranda contains 768 files each of dimension 256$\times$384$\times$384; CESM contains 61 snapshots and in total 7182 files of two types of dimensions --- 26$\times$1800$\times$3600 and 1800$\times$3600; RTM contains 3601 snapshots and each file is of dimension $449 \times 449 \times 235$.

\begin{table}[htb]
    \centering
    \caption{Basic application and dataset information}    
    \vspace{-.3cm}
    \footnotesize
    \begin{adjustbox}{width=\columnwidth}    
    \begin{tabular}{|c|p{2cm}|c|p{2cm}|}
    \hline
    \textbf{Application}&\textbf{Dataset} & \textbf{Dimensions} & \textbf{Science}\\
    \hline
    QMCPACK&einspine&33120$\times$69$\times$69& Electronic structures\\
    \hline
    RTM&3600 Snapshots&449$\times$449$\times$235& Electronic\\
    \hline
    Miranda&density, velocity, diffusity, pressure, viscosity, etc.&256$\times$384$\times$384& Hydrodynamics code for large turbulence simulations \\
    \hline
    CESM&cloud, temperature, pressure, etc.& 1800$\times$3600 & Climate\\
    \hline
    Nyx&density, temperature, etc.&512$\times$512$\times$512& Cosmology\\
    \hline
    ISABEL &temperature, speed, etc.&100$\times$500$\times$500&Weather\\
    \hline
    \end{tabular}
    \end{adjustbox}
    \label{tab:dataset}
\end{table}

We focus on SZ2 \cite{sz2}, SZ3 \cite{sz3} and their variants because our compression quality prediction method is based on the prediction-based compression model. How to estimate compression quality for transformer-based compression models is left to future work.

\subsection{Estimation of Compression Time and Ratio}

To make an estimation of compression time and ratio, we apply a decision tree regressor model on 11 features described in Section \ref{sec:predict_model}, and train on 30\% of files from each of the applications in \tablename~\ref{tab:dataset} (the remaining 70\% serves as testing data). We set 11 different error bounds from 1e-6 to 1e-1 to compress the data and collect the features for training.

\begin{figure}[htb]
    \centering
    \includegraphics[width=1\linewidth]{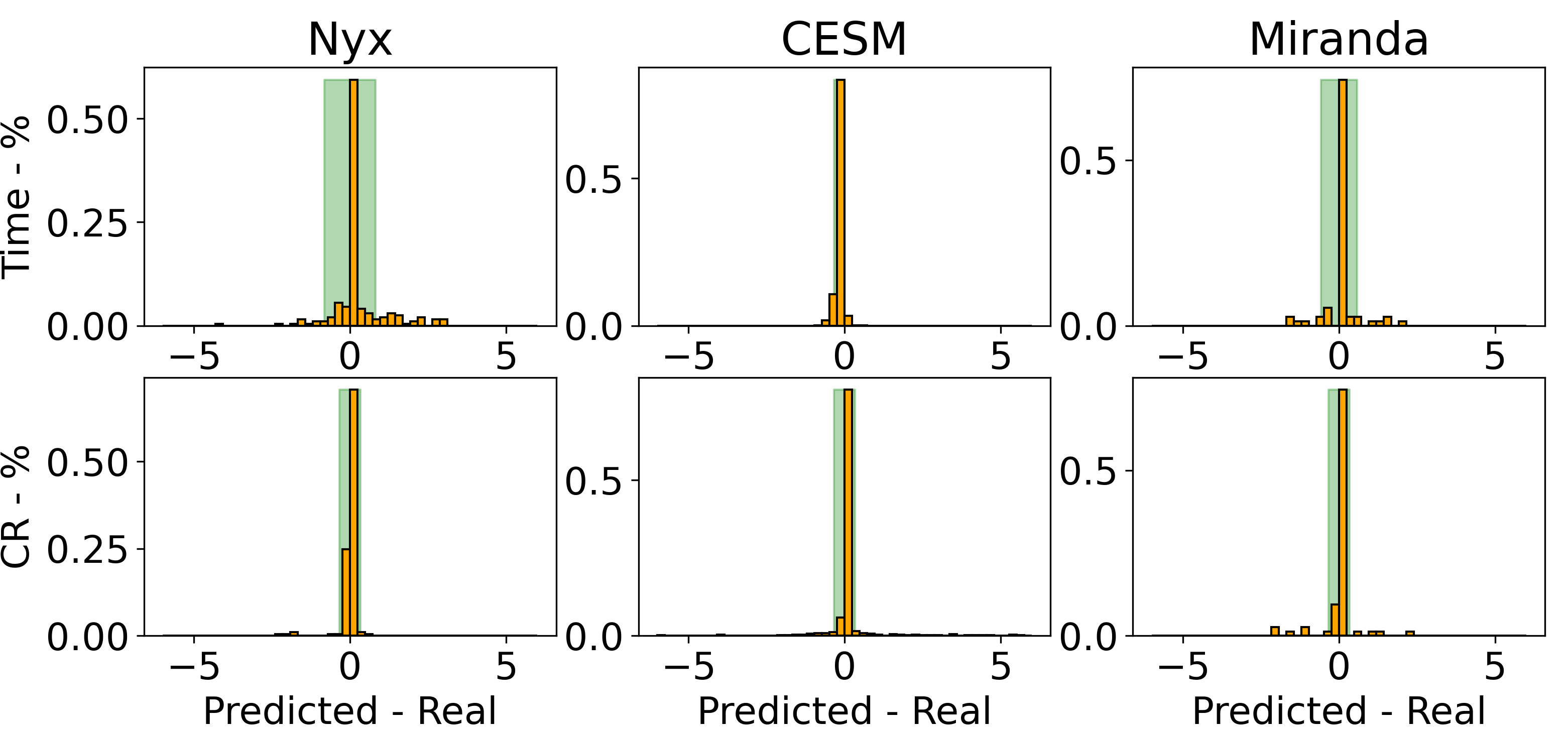}
    \vspace{-5mm}
    \caption{Nyx/CESM/Miranda application compression time and ratio prediction error distribution (measured on Bebop KNL partition): the X-axis is the difference between the predicted value and the real value, the Y-axis is the percentage for each small range of difference values.}
    \label{fig:CPTime-CR-Diff-Miranda}
\end{figure}

The distribution of the difference between the predicted values and real values is shown in  \figurename~\ref{fig:CPTime-CR-Diff-Miranda}. The green bounding box shows the 80\% confidence interval, meaning 80\% of prediction error falls into the green box. Thinner box means higher prediction accuracy. 
\figurename~\ref{fig:CPTime-CR-Diff-Miranda} indicates our prediction method performs very well, as the differences between predicted and actual values are very close to 0. 

\begin{figure}[htb]
    \centering
    \includegraphics[width=1\linewidth]{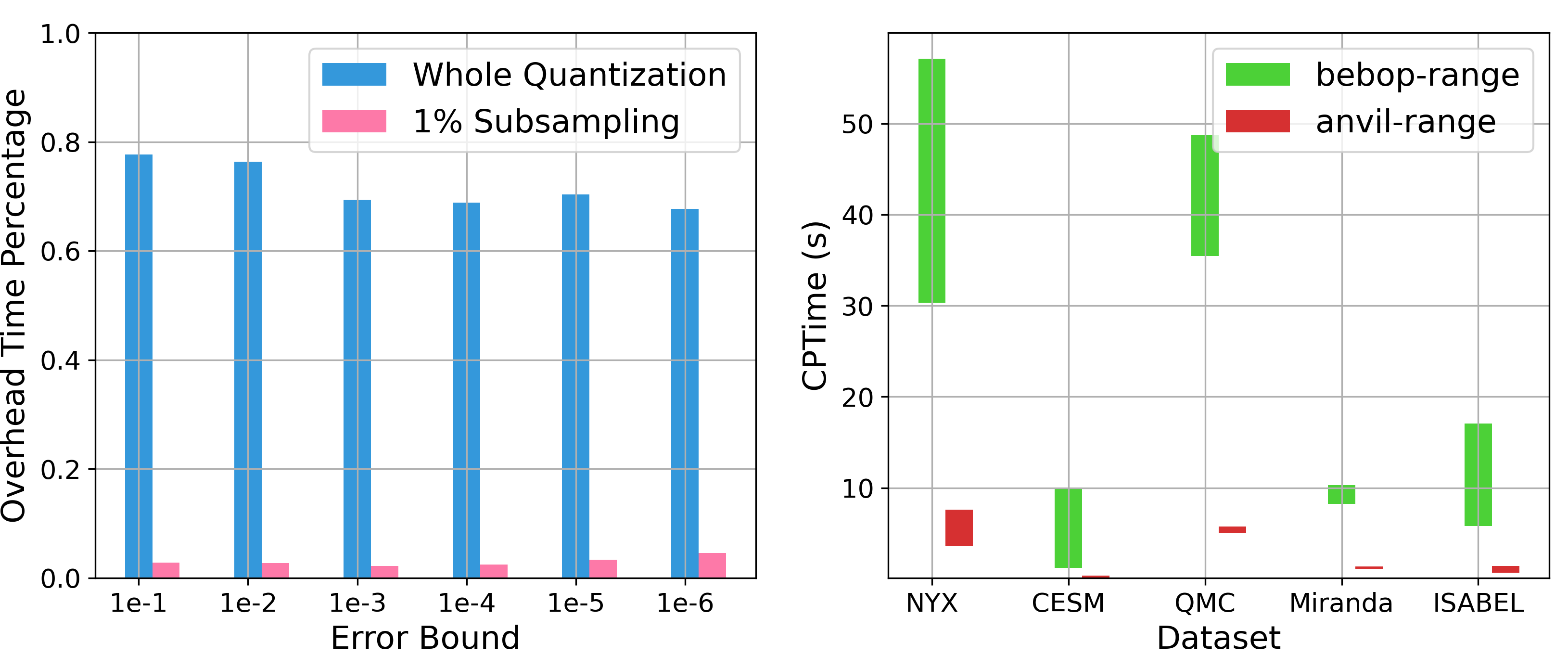}
    \vspace{-6mm}
    \caption{(A) Overhead time analysis on Nyx application; (B) Compression time range on Bebop and Anvil machines for multiple applications.}
    \label{fig:bebop-anvil-min-max}
\end{figure}

The prediction has a negligible overhead (around 1.7\%) compared with the total compression time when we sample 1\% of data (using 1 data point every 100 data points). As shown in \figurename~\ref{fig:bebop-anvil-min-max} (A), the sampling helps reduce the overhead time from more than 70\% to less than 5\%. The extracted compressor-based features $p_0$ and $P_0$ are different from the actual percentage of the zero quantization code because we run the Lorenzo prediction with the real data values instead of the reconstructed data values. 

\begin{figure}[htb]
    \centering
    \includegraphics[width=1\linewidth]{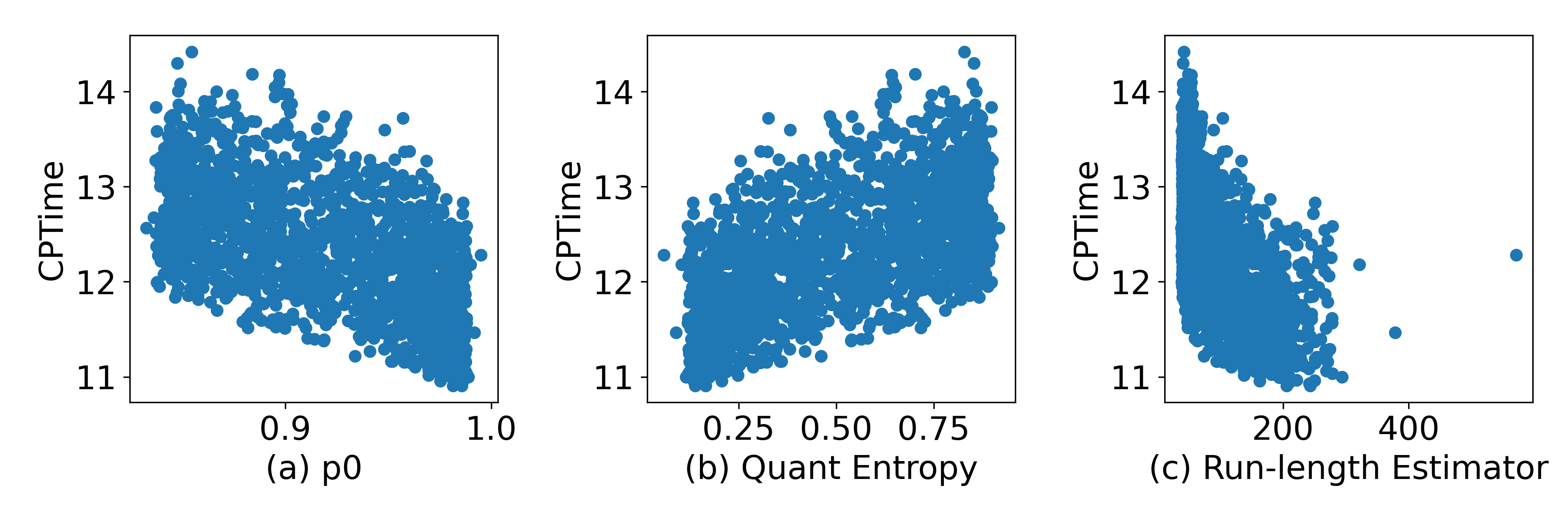}
    \vspace{-6mm}
    \caption{RTM application compression time versus compressor-level features}
    \label{fig:cptime-compresoor-features}
\end{figure}

\figurename~\ref{fig:cptime-compresoor-features} shows a high correlation between compression time and the compressor-level features. In fact, the datasets' compression times are similar with each other as long as they have the same dimensions (usually because they belong to the same application) as shown in \figurename~\ref{fig:bebop-anvil-min-max} (B). This pattern helps us estimate the overall compression time accurately in parallel compression: the rough estimation would be the number of datasets divided by the number of cores then multiplied by the average compression time per one dataset. 

\tablename~\ref{tab:prediction-examples} shows the prediction results for our datasets. We can observe from the values that the compression time is gathered into groups related to the application to which they belong. Moreover, we see that our model can always precisely predict the compression ratio and time at different error-bound settings. This is because the distribution of the quantization code changes according to error bounds, and our model captures this information with $p_0, P_0$ and the quantization entropy effectively.

\begin{table}[htb]
\centering
\caption{Compression Time and Ratio Prediction Examples: EB denotes error bound, CR denotes compression ration, CPTime denotes compression time.  P-CR and P-CPTime denote predicted compression ratio and compression time, respectively. All time-related information is measured on Bebop machine in KNL partition.}
\begin{adjustbox}{width=\columnwidth}
\begin{tabular}{|l | l || l | l|| l | l|}
\hline
\textbf{Dataset} & \textbf{EB} & \textbf{P-CR} & \textbf{CR} & \textbf{P-CPTime} & \textbf{CPTime} \\ \hline
Nyx                 & 1e-6        & 1.19          & 1.18        & 35.9              & 35.6            \\ \cline{2-6} 
Baryon Density      & 1e-4        & 3.15          & 3.10        & 32.3              & 33.3            \\ \cline{2-6} 
                    & 1e-2        & 10.40         & 10.20       & 30.3              & 30.3            \\ \hline
CESM                & 1e-6        & 1.139         & 1.135       & 1.459             & 1.456           \\ \cline{2-6} 
LHFLX               & 1e-3        & 2.56          & 2.49        & 1.97              & 1.59            \\ \cline{2-6} 
                    & 1e-2        & 5.25          & 4.43        & 1.55              & 1.50            \\ \hline
CESM                & 1e-6        & 5.36          & 6.97        & 1.61              & 1.85            \\ \cline{2-6} 
SNOWHICE            & 1e-4        & 21.0          & 21.9        & 1.55              & 1.58            \\ \cline{2-6} 
                    & 1e-3        & 48.0          & 52.8        & 1.40              & 1.48            \\ \hline
RTM-1982            & 1e-6        & 4.78          & 4.80        & 13.85             & 13.32           \\ \hline
RTM-1048            & 1e-4        & 24.72         & 24.89       & 13.1              & 13.3            \\ \hline
RTM-0594            & 1e-4        & 83.15         & 84.99       & 12.13             & 11.43           \\ \hline
Miranda             & 1e-2        & 18.99         & 16.74       & 9.57              & 9.31            \\ \cline{2-6} 
Velocity-x          & 1e-3        & 7.11          & 7.67        & 10.17             & 9.7             \\ \cline{2-6} 
                    & 1e-1        & 9.11          & 9.43        & 52.05             & 52.49           \\ \hline
\end{tabular}
\end{adjustbox}
  \label{tab:prediction-examples}
\end{table}

\subsection{Estimation of Data Quality via PSNR}

We use 50\% of gathered data for training, and perform the compression quality prediction test for the remaining 50\% of data. \tablename~\ref{tab:cesm-psnr} shows the PSNR based on 10 data files randomly selected in the CESM application, where the root mean squared error of the PSNR prediction is 13.05. \tablename~\ref{tab:ISABEL-PSNR} shows a similar prediction result for the ISABEL application, and the corresponding root mean squared error of PSNR is 14.23. Unlike the prediction of compression ratio/time which is fairly accurate, the prediction of PSNR is good in most cases yet still suffers relatively high errors occasionally on a few datasets. We plan to improve it in our future work. 

\begin{table}[htb]
\centering
\caption{Prediction of PSNR for CESM application}
\begin{adjustbox}{width=\columnwidth}
\begin{tabular}{|l|l|l|l|}
\hline
\textbf{Filename}             & \textbf{eb} & \textbf{Real PSNR} & \textbf{Predicted PSNR} \\ \hline
TMQ\_1\_1800\_3600.dat        & 1e-3        & 96.80          & 96.39              \\ \hline
CLDMED\_1\_1800\_3600.dat     & 1e-3        & 59.64          & 60.88               \\ \hline
TROP\_Z\_1\_1800\_3600.dat    & 1e-3        & 146.05         & 141.45             \\ \hline
ICEFRAC\_1\_1800\_3600.dat    & 1e-5        & 102.43         & 98.65              \\ \hline
PSL\_1\_1800\_3600.dat        & 1e-1        & 99.10          & 117.11              \\ \hline
FLNSC\_1\_1800\_3600.dat      & 1e-2        & 85.07          & 92.02              \\ \hline
ODV\_ocar2\_1\_1800\_3600.dat & 1e-5        & 79.16          & 83.92               \\ \hline
LHFLX\_1\_1800\_3600.dat      & 1e-4        & 138.92         & 136.23            \\ \hline
TREFHT\_1\_1800\_3600.dat     & 1e-3        & 99.28          & 86.82               \\ \hline
FSDTOA\_1\_1800\_3600.dat     & 1e-6        & 184.85         & 184.86              \\ \hline
\end{tabular}
\end{adjustbox}
\label{tab:cesm-psnr}
\end{table}

\begin{table}[htb]
\centering
\caption{Prediction of PSNR for ISABEL dataset}
\begin{adjustbox}{width=\columnwidth}
\begin{tabular}{|l|l|l|l|}
\hline
\textbf{Filename}        & \textbf{eb} & \textbf{Real PSNR} & \textbf{Predicted PSNR} \\ \hline
QSNOWf48\_log10.bin.dat  & 1e-2        & 72.85          & 81.11               \\ \hline
PRECIPf48\_log10.bin.dat & 1e-1        & 52.49          & 52.78               \\ \hline
QVAPORf48.bin.dat        & 1e-6        & 88.01          & 128.52              \\ \hline
PRECIPf48\_log10.bin.dat & 1e-6        & 160.18         & 160.26              \\ \hline
CLOUDf48\_log10.bin.dat  & 1e-2        & 62.07          & 88.00               \\ \hline
Wf48.bin.dat             & 1e-2        & 69.19          & 67.96               \\ \hline
QSNOWf48\_log10.bin.dat  & 1e-3        & 93.14          & 93.48               \\ \hline
CLOUDf48\_log10.bin.dat  & 1e-6        & 144.24         & 123.23              \\ \hline
Pf48.bin.dat             & 1e-2        & 98.23          & 81.21               \\ \hline
QSNOWf48\_log10.bin.dat  & 1e-6        & 160.12         & 165.35              \\ \hline
\end{tabular}
\end{adjustbox}
\label{tab:ISABEL-PSNR}
\end{table}

We explain the key reasons why the PSNR is predictable as follows. On one hand, if the quantization bins often gather around zero (especially when a relatively large error bound is used), the predicted values are likely unable to be corrected by quantization bins, leading to relatively low PSNR. On the other hand, if the zero quantization bin takes a tiny percentage, this means the quantization bins are likely very small because of the small error bounds used. In this situation, many data points would be corrected by the quantization bins or stored as they are based on the SZ compression model, thus leading to relatively high PSNR. 

We explain why the prediction of PSNR may not be as precise as the compression ratio's prediction as follows. In fact, when $p_0$ and the quantization entropy are in the middle, most data points can still be the quantization-based reconstructed data, and it is unclear how far away these data points are from the original data values. They can be either an error bound away or quite close, therefore it is unclear how they will contribute to the final PSNR based on the selected features.

\begin{figure}[htb]
    \centering
    \includegraphics[width=1\linewidth]{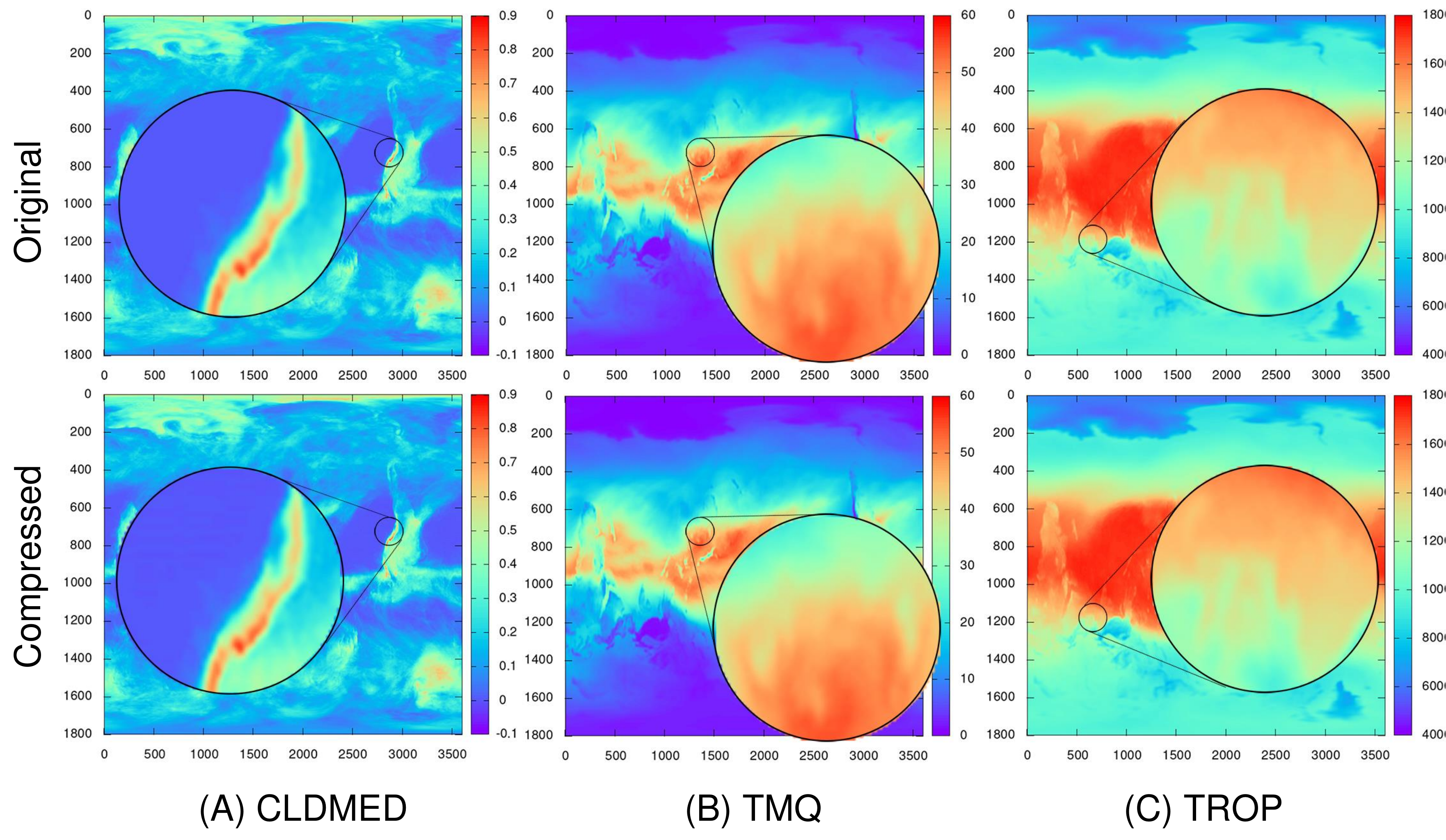}
    \vspace{-6mm}
    \caption{CESM data visualization comparison between original and compressed data: The PSNRs are 59.64, 96.80, and 146.05 respectively, and there is no obvious visual difference between the original and compressed data.}
    \label{fig:CESM-visuals}
\end{figure}

With the settings shown in \tablename~\ref{tab:cesm-psnr}, we visualize the original and compressed data of three data files in \figurename~\ref{fig:CESM-visuals}. From our experience, when PSNR is higher than 50, there is no visible visual difference between the original and compressed data. Therefore, when the predicted PSNR is high, we are confident that the compressed data will be of a good quality for post-analysis.

\subsection{Transfer Datasets with Parallel Compression}

\begin{table*}[htb]
\centering
\caption{Data Transfer Test Among Purdue Anvil, Argonne Bebop, and NERSC Cori: T/Speed(NP) is the transfer time/effective speed without compression; T/Speed(CP) is with compression, while each file has its own compressed file; T/Speed(OP) is compression with our file grouping optimization. The CPTime is the total compression time before the transfer begins, and DPTime is the total decompression time after the files are transferred. Total T is the total time using our solution, including compression, transfer, and decompression time. Gain is the performance improvement calculated by (T(NP) - Total T)/ T(NP)}.
\vspace{-6mm}
\begin{adjustbox}{width=\linewidth}
\begin{tabular}{|l|l|l|l|l|l|l|l|l|l|l|r|r|}
\hline
Dataset & \# Files & Direction                 & T(NP) & Speed(NP) & T(CP)       & Speed(CP) & T(OP)        & Speed(OP) & CPTime & DPTime & \multicolumn{1}{l|}{Total T} & \multicolumn{1}{l|}{Reduced} \\ \hline
        &          & Anvil-\textgreater{}Cori  & 446s   & 3.63GB/s  & 87s          & 2.55GB/s  & \textbf{75s}  & 2.93GB/s  & 32.48s & 68.7s  & 176.18s                      & 60\%                      \\ \cline{3-13} 
CESM    & 7182     & Anvil-\textgreater{}Bebop & 1685s  & 960MB/s   & 269s         & 822MB/s   & \textbf{250s} & 885MB/s   & 32.54s & 126s   & 408.54s                      & 76\%                      \\ \cline{3-13} 
1.61TB  &          & Bebop-\textgreater{}Cori  & 1484s  & 1.09GB/s  & 268s         & 827MB/s   & \textbf{217s} & 1.02GB/s  & 135s   & 69.4s  & 421.4s                       & 72\%                      \\ \hline
        &          & Anvil-\textgreater{}Cori  & 181s   & 3.76GB/s  & 15s          & 932MB/s   & \textbf{11s}  & 1.27GB/s  & 8.99s  & 21.8s  & 41.79s                       & 77\%                      \\ \cline{3-13} 
RTM     & 3601     & Anvil-\textgreater{}Bebop & 784s   & 870MB/s   & 28s          & 503MB/s   & \textbf{20s}  & 712MB/s   & 9.03s  & 41.2s  & 70.23s                       & 91\%                      \\ \cline{3-13} 
682GB   &          & Bebop-\textgreater{}Cori  & 623s   & 1.09GB/s  & 25s          & 544MB/s   & \textbf{18s}  & 795MB/s   & 56s    & 21.9s  & 95.9s                        & 85\%                      \\ \hline
        &          & Anvil-\textgreater{}Cori  & 35s    & 3.32GB/s  & \textbf{11s} & 1.22GB/s  & 13s           & 974MB/s   & 6.52s  & 3.07s  & 20.59s                       & 41\%                      \\ \cline{3-13} 
Miranda & 768      & Anvil-\textgreater{}Bebop & 134s   & 870MB/s   & \textbf{23s} & 577MB/s   & 30s           & 444MB/s   & 6.27s  & 8.89s  & 38.16s                       & 72\%                      \\ \cline{3-13} 
115GB   &          & Bebop-\textgreater{}Cori  & 119s   & 972MB/s   & \textbf{19s} & 676MB/s   & 24s           & 553MB/s   & 8.83s  & 3.08s  & 30.91s                       & 74\%                      \\ \hline
\end{tabular}
\end{adjustbox}
\label{tab:data-transfer}
\end{table*}

We now investigate the overall transfer performance when utilizing parallel compression on supercomputers. 
We use three applications CESM, RTM, and Miranda to analyze parallel compression performance.

\begin{figure}[htb]
    \centering
    \includegraphics[width=1\linewidth]{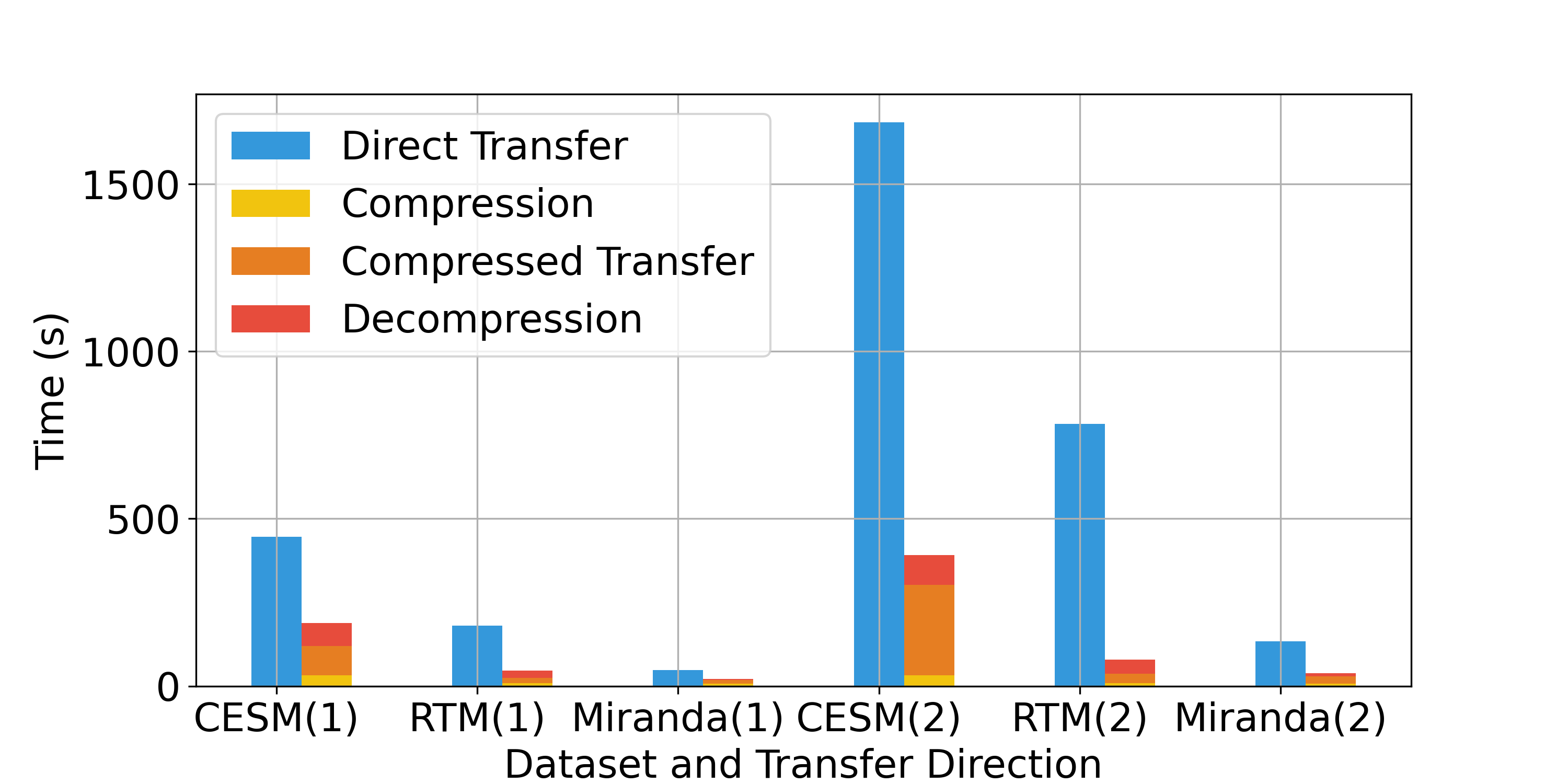}
    \vspace{-6mm}
    \caption{Transfer time comparison between direct transfer and transfer with compression. (1) means transferring from Purdue Anvil to NERSC Cori, (2) means transferring from Purdue Anvil to Argonne Bebop.}
    \label{fig:transfer-comparison}
\end{figure}

\figurename~\ref{fig:transfer-comparison} shows the time reduction because of our parallel compression applied in the data transfer. The compression time is measured on the Purdue Anvil machine with 16 nodes (each node uses 128 CPU cores, and in total 2048 CPU cores), while decompression is measured on Bebop for experiment (1) and on Cori for experiment (2) with 8 nodes (each node has 32 CPU cores and in total 256 CPU cores). We see increased transfer speed when using parallel compression because (1) the total file size is much smaller and (2) the compression time is minimized by parallelization. The node waiting time on Purdue Anvil machine is  negligible in our experiments because compression tasks can immediately be scheduled. On Bebop and Cori, however, the node waiting time varies. When there were idle nodes, the waiting time was between 0s to 30s, but sometimes it took a few minutes or even hours to get an available compute node. The behavior is highly dependent on other users' tasks, and we could not conclude any quantifiable patterns on the expected node waiting time. Our sentinel program can ensure the worst case would be transferring the data without compression.


\tablename~\ref{tab:data-transfer} shows the comparison between direct transfer without compression and transfer involving our parallel compression method.
Because of a significant reduction in file sizes, we can see an obvious reduction in the transfer time for all three applications. We notice that the effective transfer speed drops after compression without file grouping. This is because the files are smaller while the number of directories and the number of files stays constant. This result aligns with the pattern shown in \tablename~\ref{tab:file-transfer-pattern}. Because large files generally transfer faster in the network than small files, our file grouping strategy helps counter the speed reduction for the RTM and CESM applications. For the Miranda application, the grouped files do not transfer faster because, after grouping, there are only 8 files and it has not reached the number of concurrent threads available in the Globus Transfer Service. This result also shows that we should strategically group files into multiple groups instead of simply connecting all compressed files into one large file. Moreover, making all cores write to the same file would cause I/O contention and add overhead to the file grouping process.
\section{Conclusion and Future Work}
\label{sec:conclusion}

We developed a novel data transfer framework, Ocelot, that integrates Globus transfer with transparent error-bounded prediction-based lossy compression. We proposed a model to predict compression ratio/time and data quality for user defined compression settings with little overhead. Based on our evaluation on six real-world scientific datasets, we report the following key findings.

\begin{itemize}
    \item Compression time/ratio and PSNR are predictable by using various categories of features. By doing 1\% sampling, we can reduce the overhead required to finish the prediction to 1.7\% of compression time---a small cost when compared with transfer time.
    
    \item Scientific data transfer performance can be greatly improved by applying parallel compression. We use FuncX to further control the node waiting time on supercomputers, and minimize the transfer time for given datasets. More than 90\% of the transfer time can be reduced by this method.
    
    \item Network transfer speed can be significantly affected by file size and number of files. A few large files generally transfer faster than many small files. We can improve transfer speed by grouping smaller compressed files, and the transfer time can be reduced by more than 25\%  because of file grouping.
    
    \item
    While the use of more CPU cores can improve compression and decompression performance, I/O contention can become a problem in the decompression case. It is generally better to use more CPU cores for compression and fewer CPU cores for decompression.
\end{itemize}


We selected features that were simple to derive and fast to train and make predictions, but there is still room to extract better features to improve  prediction accuracy. In addition, our model requires seeing the dataset in advance to make predictions and has very limited generalization to other datasets. Moreover, we lack effective time/ratio prediction methods for transformer-based compressors like ZFP\cite{zfp} and TTHRESH\cite{tthreash}. In the future, we will look into other features, particularly those that do not require processing of the data, to see if we can make accurate predictions on datasets that have never appeared in the training set. We will also investigate additional compressor types and work to identify features that are suitable for transformer-based compressors.


\section*{Acknowledgments}
The material was supported by the U.S. Department of Energy, Office of Science, Advanced Scientific Computing Research (ASCR), under contract DE-AC02-06CH11357, and supported by the National Science Foundation under Grant OAC-2003709 and OAC-2104023. We acknowledge the computing resources provided on Bebop (operated by Laboratory Computing Resource Center at Argonne).

\bibliographystyle{IEEEtran}
\bibliography{reference}

\end{document}